\documentclass[12pt]{article}
\usepackage{epsf,epsfig,amssymb,amsmath,graphicx,float}
\usepackage{color}

\begin{document}
\renewcommand{\thefootnote}{\fnsymbol{footnote}}

\begin{titlepage}

\begin{center}

\vspace{1cm}

{\Large {\bf Asymmetric Dark Matter and the Scalar--Tensor Model }}

\vspace{1cm}

{\bf Shun-Zhi Wang}$^a$\footnote{wsz08@pku.edu.cn},
{\bf Hoernisa Iminniyaz}$^b$\footnote{Corresponding author, wrns@xju.edu.cn},
{\bf Mamatrishat Mamat}$^c$\footnote{mmtrxt@xju.edu.cn},
\vskip 0.15in
{\it
$^a${College of Fundamental Studies, Shanghai University of Engineering Science ,\\ Shanghai 201620, P. R. China}\\
$^{b,c}${School of Physics Science and Technology, Xinjiang University, \\
Urumqi 830046, P. R. China} \\
}

\abstract{The relic abundance of asymmetric Dark Matter
  particles in the scalar--tensor model is analyzed in this article.
  We extend the numeric and analytic calculation of the relic density of
  the asymmetric Dark Matter in the standard cosmological scenario to the
  nonstandard cosmological scenario. We focus on the scalar--tensor model.
  Hubble expansion rate is changed in the nonstandard cosmological scenario.
  This leaves its imprint on the relic density of Dark Matter particles. In this article we investigate to what extent the asymmetric Dark Matter particle's relic density is changed in the scalar--tensor model. We use the observed present day Dark Matter abundance to find the constraints on
  the parameter space in this model. }
\end{center}
\end{titlepage}
\setcounter{footnote}{0}

\section{Introduction}

The fact that the Universe contains a large amount of non-baryonic
Dark Matter is well established by astronomical
observations\cite{Zwicky:1937zza,Bertone:2004pz, wmap}.
The Dark Matter has some extraordinary properties, \emph{e.g.},
non--luminous and non--absorbing.
Among the various Dark Matter candidate particles, ``Weakly
Interacting Massive Particles (WIMPs)'' are considered to be best
motivated candidates since the present Dark Matter relic density can
be explained in this scenario naturally \cite{PDG}. WIMPs are
expected to interact among themselves and with the ordinary matter
only through the gravity and weak interactions, with masses between
10 GeV to a few TeV. It is assumed that when the universe is
radiation dominated, WIMPs are in thermal and chemical
equilibrium with the rest of the particles.
 As the Universe expands and the temperature of the universe
drops, the Dark Matter particles decouple from the equilibrium and
resulted in essentially constant co--moving density
(``freeze--out''). If the entropy of the radiation and matter is
conserved during the radiation--dominated era in which the Dark
matter particle is produced, the relic abundance of Dark Matter
particles can be calculated by solving Boltzmann equation in the
standard method. It is shown that for appropriate WIMP annihilation
cross section, the resulting relic abundance of Dark Matter is of
the same order of magnitude as the observed value.

Most of the WIMPs under consideration are symmetric Dark Matter
particles whose particle and anti--particle are the same. The well
known example is Neutralino which is Majorana particle appeared in
supersymmetric model with conserved $R$--parity \cite{review}.
However, the status of experiment detection of Dark Matter urges
people to broaden the search scope. In fact, the idea of asymmetric
Dark Matter has been proposed as early as
1980's\cite{adm-models,frandsen}. In asymmetric Dark Matter
scenario, Dark Matter particles and antiparticles are not identical
and have an asymmetry similar to the baryonic asymmetry.
Furthermore, it is expected that both asymmetries may have the
common origin appeared in the early Universe.

The Dark Matter relic density depends on the characteristics of the
Universe, especially the expansion rate of the Universe before
Big Bang Nucleosynthesis \cite{Gelmini:2006pw}. The expansion rate
may be changed through additional contributions to the total energy
density\cite{Salati:2002md}, anisotropic expansion
\cite{Kamionkowski} and a modification of general relativity
\cite{Kamionkowski,Catena}. The relic abundances of asymmetric Dark
Matter particles in the standard cosmological scenario are
investigated in papers \cite{Iminniyaz:2011yp, GSV}.
The asymmetric Dark Matter relic density in some nonstandard cosmological
scenarios has been discussed \cite{Iminniyaz:2013cla, Gelmini:2013awa}.
In scalar--tensor model, it is shown that
the relic density of Dark Matter is increased with the changing
expansion rate of the Universe\cite{Catena:2004ba}.
Obviously, more detailed investigation is needed for the relic
density of asymmetric Dark Matter in the nonstandard cosmological
scenarios to find the effects of the modification of the Hubble rate
on asymmetric Dark Matter.

In this paper, the discussion about the relic density of asymmetric
WIMP Dark Matter in the standard cosmological scenario is extended
to scalar--tenser model. We are not concerned about the mechanism
for the asymmetry and assume that there are more particles than the
anti--particles in the beginning. We find that the enhanced Hubble
rate in scalar-tensor model changes the relic density of both
particles and anti--particles. When the cross section is not large,
the deviation of the relic density of particle and anti--particle in
scalar--tensor model is not large. They are almost in the same
amount at the final temperature for smaller cross section in this
model. For large cross section, the relic density of anti--particle
is depressed and there are only particles left at the final
temperature. In the end, the observed present day Dark Matter
abundance is applied to constrain the parameter space in the
scalar--tensor model.

This paper is arranged as follows. In section 2, the relic density
of asymmetric Dark Matter in scalar--tensor model is discussed.
In section 3, we constrained the
expansion rate in scalar--tensor model using the observed Dark Matter
abundance. The conclusions and discussions are given in section 4.

\section{Relic Abundance of Asymmetric Dark Matter in the Scalar--Tensor Model}

Nonstandard cosmological scenarios are based on the alternative theories of
gravity rather than the General Relativity. The scalar--tensor theories are
the well studied alternative theories of gravity. In scalar--tensor model,
the gravity sector is extended to contain an extra scalar field $\phi$ which
couples to the matter fields through the metric tensor. It is shown that
in such models with a single matter sector there is an enhancement of $H$
before BBN~\cite{Catena:2004ba,Santiago:1998ae}. Typically the expansion rate
$H$ in these models is given by~\cite{Catena:2004ba}:
\begin{eqnarray}\label{Hratio}
      H & = & f_\phi(T) H_{\rm std},
\end{eqnarray}
where $f_\phi(T)\simeq 2.19\times 10^{14}\left({T_0}/{T}\right)^{0.82}$ is an
enhancement factor and $T_0$ is the present temperature of the Universe. In
terms of new variable $x = m_{\chi}/T$ where $m_\chi$ is the mass of the Dark
Matter particles $\chi$ and $\bar{\chi}$, this factor is expressed as
$f_\phi(x)\simeq 9.65\times 10^3\left({\rm GeV}/m_\chi\right)^{0.82}x^{0.82}$.
Here $ H_{\rm{std}} = \sqrt{8\pi^3 g_*/90} T^2/M_{\rm P}\equiv
H(m_{\chi})/x^2$
with $ H(m_{\chi}) = \sqrt{8\pi^3 g_*/90} m_{\chi}^2/M_{\rm P} $, $g_*$
is the effective number of the relativistic degrees of freedom.
At the low temperature $T_{\rm tr}$, the factor $f_\phi(T)$ must tend to 1
so that the standard cosmology resumes \cite{Catena:2004ba}. $T_{\rm tr}$ is
called the transition temperature at which point the Hubble expansion rate is
returned to the standard Hubble rate.

The abundances of asymmetric Dark Matter $\chi$ and $\bar{\chi}$ are
characterized by the number densities $n_{\chi}$, $n_{\bar\chi}$,
respectively, here $\bar\chi \neq \chi$. The relic densities of $\chi$ and
$\bar\chi$ are obtained by solving their Boltzmann equations respectively.
The Boltzmann equations describe the time evolution of the number
densities $n_{\chi}$,
$n_{\bar\chi}$ in the expanding universe. Based on our assumption that only $\chi \bar \chi$ pairs can
annihilate into Standard Model (SM) particles, while $\chi\chi$ and
$\bar \chi \bar\chi$ pairs can not, the Boltzmann equations are:
\begin{eqnarray} \label{eq:boltzmann_n}
\frac{{\rm d}n_{\chi}}{{\rm d}t} + 3 H n_{\chi} &=&  - \langle
\sigma v\rangle
  (n_{\chi} n_{\bar\chi} - n_{\chi,{\rm eq}} n_{\bar\chi,{\rm eq}})\,;
  \nonumber \\
\frac{{\rm d}n_{\bar\chi}}{{\rm d}t} + 3 H n_{\bar\chi} &=&
   - \langle \sigma v\rangle (n_{\chi} n_{\bar\chi} - n_{\chi,{\rm
       eq}} n_{\bar\chi,{\rm eq}})\,,
\end{eqnarray}
where $\langle \sigma v \rangle$ is
the thermally averaged annihilation cross section times relative velocity of
the annihilating particles.
$ n_{\chi,{\rm eq}} = g_\chi \left( m_\chi T/2 \pi \right)^{3/2}
                  {\rm e}^{(-m_\chi + \mu_\chi)/T} $ and
$n_{\bar\chi,{\rm eq}} = g_\chi {\left( m_\chi T/2 \pi \right)}^{3/2}
                  {\rm e}^{(-m_\chi - \mu_{\chi})/T} $ are the
eqilibrium number density for particle and anti--particle, where $g_{\chi}$ is
the number of the internal degrees of freedom of $\chi$ and $\bar{\chi}$
particles. In equilibrium state, the chemical potential of the particles and
anti--particles are assumed to be $\mu_{\bar\chi} = -\mu_\chi$.

Usually it is assumed that at high temperature, $\chi$ and $\bar\chi$ particles
are in thermal equilibrium in the early universe.
Following the expansion of the universe, the
temperature drops down. The number densities $n_{\chi,{\rm eq}}$ and
$n_{\bar\chi,{\rm eq}}$ decrease exponentially for $m_\chi > |\mu_\chi|$ when
$T < m_\chi$. When the interaction rates
$\Gamma = n_{\chi} \langle \sigma v \rangle$ for
particle and $\bar{\Gamma} = n_{\bar\chi} \langle \sigma v \rangle$ for
anti--particle drop below the Hubble expansion rate $H$, $\chi$ and
$\bar\chi$ particles decouple from chemical equilibrium and their co--moving
number densities are fixed. The temperature at this decoupling point is called
the ``freeze--out" temperature.

Next we reformulate the Boltzmann equations (\ref{eq:boltzmann_n})
in terms of the dimensionless quantities $Y_\chi =n_\chi/s$,
$Y_{\bar\chi} = n_{\bar\chi}/s$ for convenience. The entropy
density $s$ is given by $s =(2\pi^2/45)g_*T^3 $.
If we furthermore assume ${\rm d}{g_*}/{\rm d}x \simeq
0$ during the radiation--dominated period, the Boltzmann
equations (\ref{eq:boltzmann_n}) become
\begin{equation} \label{eq:boltzmann_Y}
\frac{d Y_{\chi}}{dx} = - \frac{\lambda \langle \sigma v
\rangle}{x^2 f_\phi(x)}~ (Y_{\chi}~ Y_{\bar\chi} - Y_{\chi, {\rm
eq}}~Y_{\bar\chi, {\rm eq}}   )\,;
\end{equation}
\begin{equation} \label{eq:boltzmann_Ybar}
\frac{d Y_{\bar{\chi}}}{dx} = - \frac{\lambda \langle \sigma v
\rangle}{x^2 f_\phi(x) }~
 (Y_{\chi}~Y_{\bar\chi} - Y_{\chi, {\rm eq}}~Y_{\bar\chi, {\rm eq}} )\,,
\end{equation}
where we have introduced the constant
%
\begin{equation} \label{lambda}
\lambda =\frac{x^3s}{H(m_{\chi})}\,
        =\sqrt{\frac{\pi}{45}g_*}\,m_{\chi} M_{\rm P}\,.
\end{equation}
From Eqs.(\ref{eq:boltzmann_Y}), (\ref{eq:boltzmann_Ybar}) we obtain
%
%
\begin{equation}  \label{eq:c}
 Y_{\chi} - Y_{\bar\chi} = C\,,
\end{equation}
where $C$ is a constant. This means that the difference of the co--moving
densities of
the particles and anti--particles is conserved. Considering the
relation (\ref{eq:c}), the Boltzmann equations
(\ref{eq:boltzmann_Y}) and (\ref{eq:boltzmann_Ybar}) become
\begin{equation} \label{eq:Yc}
\frac{d Y_{\chi}}{dx} = - \frac{\lambda \langle \sigma v
\rangle}{x^2 f_\phi(x)}~
      (Y_{\chi}^2 - C Y_{\chi} - P     )\, ;
\end{equation}
\begin{equation} \label{eq:Ycbar}
\frac{d Y_{\bar{\chi}}}{dx} = - \frac{\lambda \langle \sigma v
\rangle}{x^2 f_\phi(x)}~
 (Y_{\bar\chi}^2 + C Y_{\bar\chi}  - P)\,,
\end{equation}
where $P = Y_{\chi,{\rm eq}} Y_{\bar\chi,{\rm eq}} = (0.145
g_{\chi}/g_*)^2\,x^3\,{\rm e}^{-2x}$

\begin{figure}[h!]
  \begin{center}
    \hspace*{-0.5cm} \includegraphics*[width=9cm]{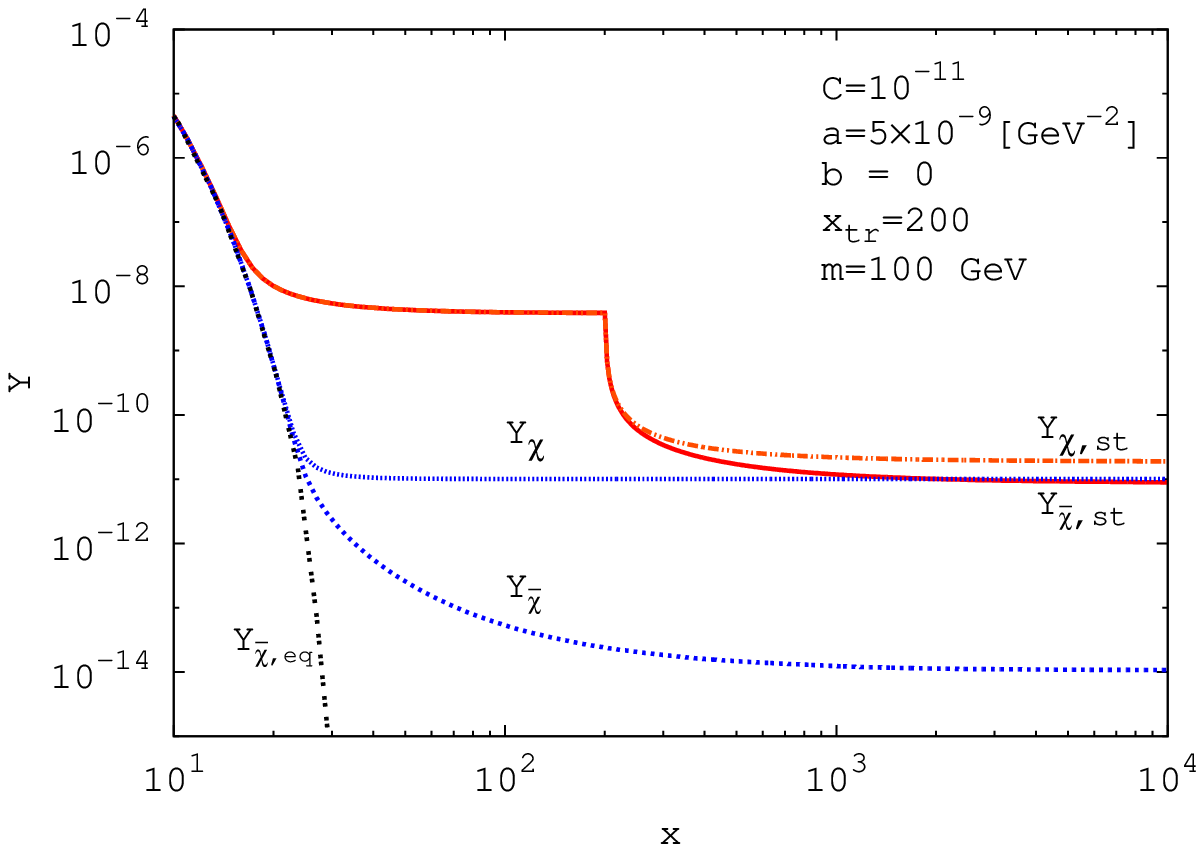}
    \put(-115,-12){(a)}
    \hspace*{-0.5cm} \includegraphics*[width=9cm]{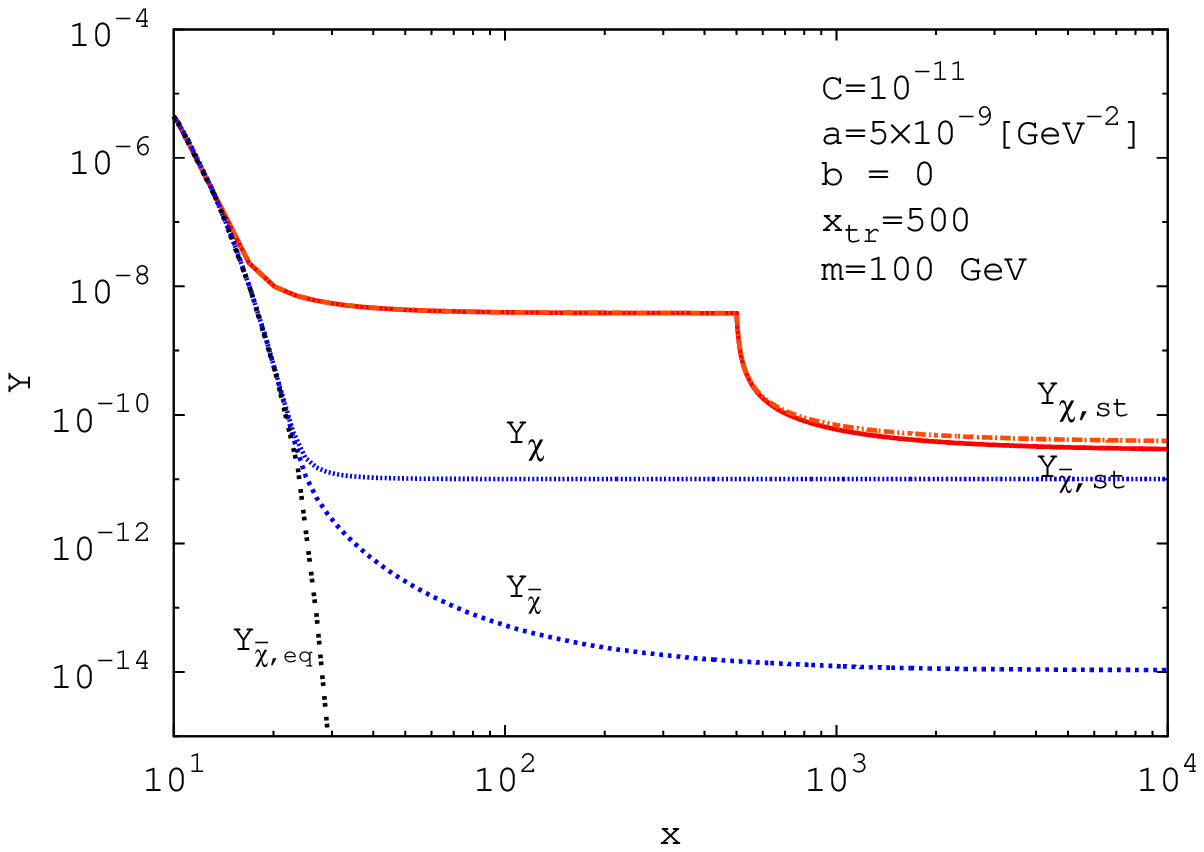}
    \put(-115,-12){(b)}
    \vspace{0.5cm}
   \hspace*{-0.5cm} \includegraphics*[width=9cm]{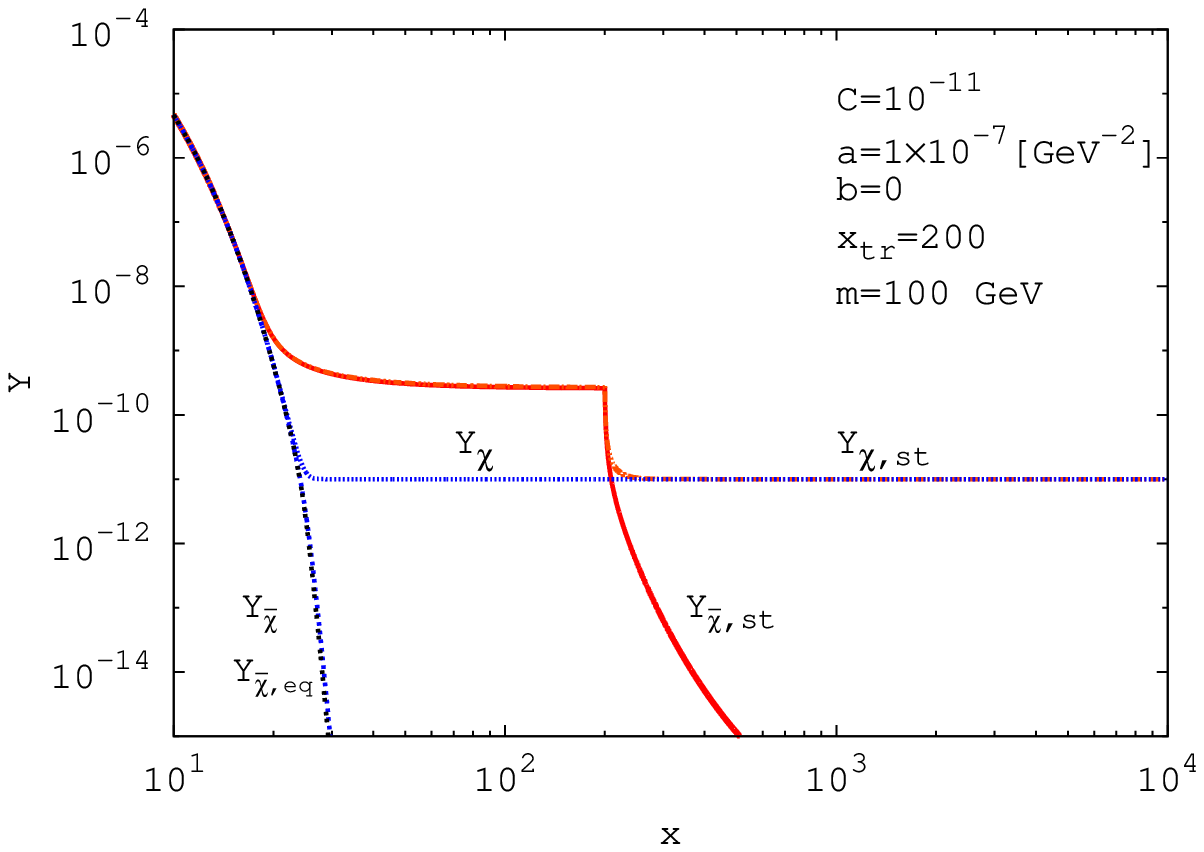}
    \put(-115,-12){(c)}
    \hspace*{-0.5cm} \includegraphics*[width=9cm]{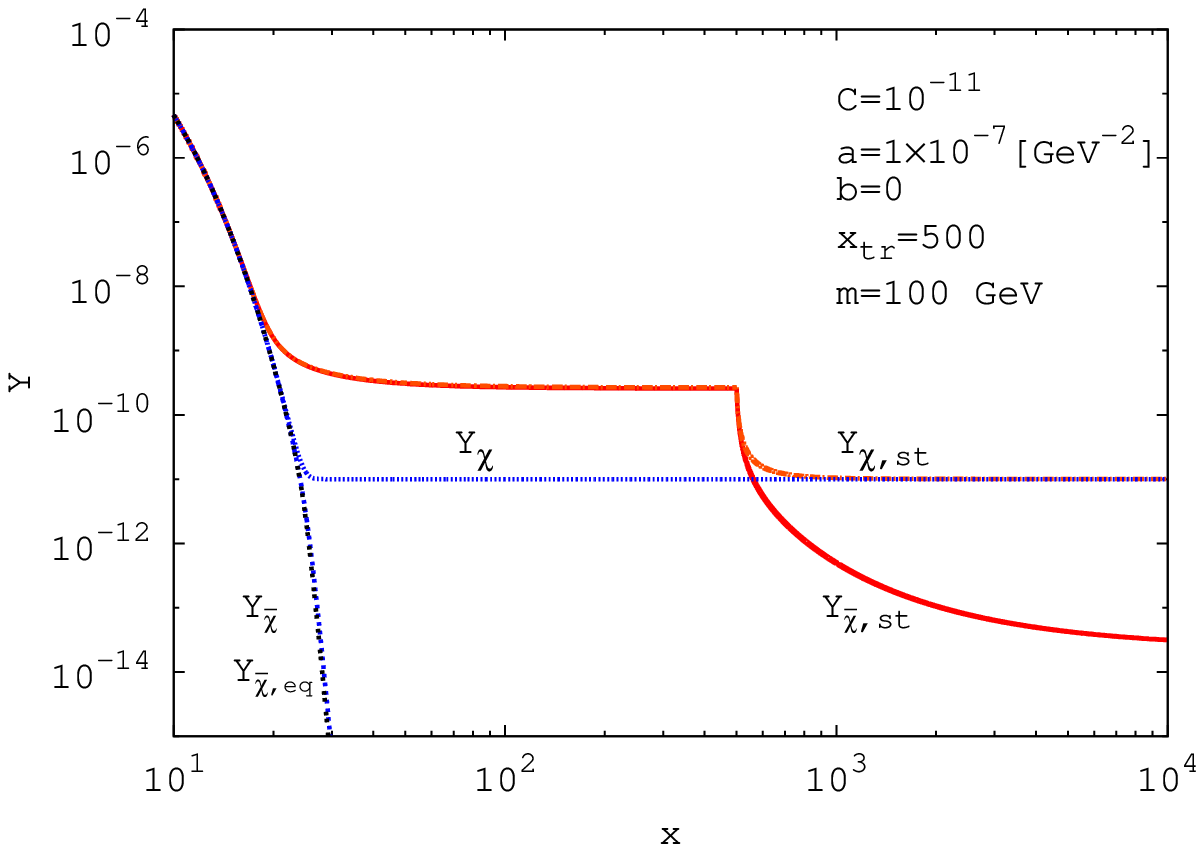}
    \put(-115,-12){(d)}
     \caption{\label{fig:a}
     \footnotesize The relic abundance $Y$ as a function of the
     inverse--scaled temperature $x=m_\chi/T$ for s--wave annihilation (b=0).
     Here $Y_{\chi, {\rm st}}$ and $Y_{\bar\chi,{\rm st}}$ are the abundance
     for particle $\chi$ and anti--particle $\bar\chi$ in the scalar--tensor
     model, while $Y_{\chi}$ and $Y_{\bar\chi}$  are for the standard
     cosmological scenario. Panel (a), (c) are for $x_{\rm{tr}}=200$ and panel
     (b), (d) are for $x_{\rm{tr}}=500$. }
  \end{center}
\end{figure}
Eqs.(\ref{eq:Yc}), (\ref{eq:Ycbar}) can be solved numerically. The
relic abundance for particle and anti--particle $Y$ as a function of
the inverse--scaled temperature $x=m_\chi/T$ for s--wave annihilation is
given numerically in Fig.\ref{fig:a}. The double dotted (black) line is the
equilibrium
value of the anti--particle abundance. The solid (red) line is the
relic abundance for anti--particle and dot--dashed (red) line is
that for the particle in the scalar--tensor model. For comparison,
we also give the relic abundance of Dark Matter in the standard
cosmological scenario where the dotted (blue) line is the relic
abundance for the anti--particle while the short dashed (blue) line
for particle. The calculations are carried out for different cross
sections and different inverse--scaled transition temperatures
$x_{\rm{tr}}=m_\chi/T_{\rm{tr}}$. In the calculations we take the thermally averaged cross section as $
   \langle \sigma v \rangle
      \approx a + 6\,b x^{-1} + {\cal O}(x^{-2})\,, $
where $a$ is dominant for s--wave $\chi \bar{\chi}$ annihilation, while $b$ for p--wave annihilation. In Fig.\ref{fig:a} we take $ a = 5.0 \times 10^{-9}$
GeV$^{-2}$, $b = 0$ for panels (a), (b) and  $ a = 1 \times 10^{-7}$
GeV$^{-2}$, $b = 0$ for (c), (d). In panels (a), (c) we consider
$x_{\rm{tr}}=200$ and in panels (b), (d) $x_{\rm{tr}}=500$. In all
cases, we take $C = 10^{-11}$, $m_\chi = 100$ GeV, $g_\chi=2$ and
$g_*=90$.

Two common features of the relic abundances in scalar-tensor model
can be shown in Fig.\ref{fig:a}. First, when the inverse--scaled temperature is
larger than the inverse--scaled transition temperature the particle and
anti--particle abundance are kept in the same amount which are larger than
the standard scenario. This is because of
the enhancement of the expansion rate in scalar-tensor model. Due to
this enhancement, the asymmetric Dark Matter particles freeze--out earlier
than the standard case, leading to the increases of particle and
anti--particle abundances. For example, the particle abundance is
increased as large as $9 \times 10^{-9}$ in (a) of Fig.\ref{fig:a}. Because
$ Y_{\chi} - Y_{\bar{\chi}} = C$, comparing
to asymmetry factor $C = 10^{-11}$, this value is so large that the
difference between particle and anti--particle abundance can not be
manifested in the figure. Thus it seems that the particle and
anti--particle are in the same amount before transition temperature. Second, sudden decreases 
of the relic abundances in scalar-tensor model appear at the transition
temperatures $x_{\rm tr}$ in Fig.\ref{fig:a}.
 This corresponds to the behavior of the Hubble parameter in the scalar-tensor model around
 the transition temperatures $T_{\rm tr}$. For transition temperatures varying
 from 1 MeV to a few GeV, the Hubble parameter decreases suddenly by a few
orders of magnitude at $T_{\rm tr}$.

Once the temperature drops below the transition temperature, the
Hubble expansion rate is resumed to the standard Hubble rate, and
the deviation of the abundance for the particle and
anti--particle is appeared. The behavior of the relic abundance is
influenced by the cross sections. For smaller s--wave annihilation
cross section $ a = 5 \times 10^{-9}$ GeV$^{-2}$, the relic
abundances $Y_{\chi, {\rm st}}$ and $Y_{\bar\chi,{\rm st}}$ for
particle $\chi$ and anti--particle $\bar\chi$ in scalar--tensor
model are greater than that of the standard case in Fig.\ref{fig:a}
for both panel (a) and panel (b). When the cross section is larger
as $ a = 1 \times 10^{-7}$ GeV$^{-2}$, the final abundance for
particle in scalar--tensor model is same with the standard case;  while the anti--particle
abundance in scalar--tensor model is decreased sharply as the
temperature drops for
the transition temperature $x_{\rm tr} = 200 $. The decrease is smaller for the transition
temperature $x_{\rm tr} = 500 $. Notice that for the cross section
$a = 1\times 10^{-7}$ GeV $^{-2}$ the relic abundance of
anti-particle in standard cosmology $Y_{\bar{\chi}}$ are also
same as the equilibrium value $Y_{\bar{\chi}, \rm eq}$.

In the following, we give the approximate analytic solutions of the relic densities for asymmetric Dark Matter in scalar-tensor model.
The procedure is the same as in standard cosmological scenario \cite{Iminniyaz:2011yp}.
For convenience, we introduce the quantity
$\Delta_{\bar\chi} = Y_{\bar\chi} - Y_{\bar\chi,{\rm eq}}$.
Using $\Delta_{\bar\chi}$, the Boltzmann equation (\ref{eq:Ycbar}) can
be rewritten as:
\begin{equation} \label{eq:delta}
\frac{d \Delta_{\bar\chi}}{dx} = - \frac{d Y_{\bar\chi,{\rm
eq}}}{dx} - \frac{\lambda \langle \sigma v \rangle}{x^2 f_\phi(x)}~
\left[\Delta_{\bar\chi}(\Delta_{\bar\chi} + 2 Y_{\bar\chi,{\rm eq}})
      + C \Delta_{\bar\chi}   \right]\, .
\end{equation}
At sufficiently high temperature, $Y_{\bar\chi}$ tracks its
equilibrium value $Y_{\bar\chi,{\rm eq}}$ very closely. In that
regime the quantity $\Delta_{\bar\chi}$ is small, and $d
\Delta_{\bar\chi}/dx$ and $\Delta_{\bar\chi}^2$ are negligible. The
Boltzmann equation (\ref{eq:delta}) then becomes
\begin{equation}\label{eq:delta_simp}
     \frac{d Y_{\bar\chi,{\rm eq}}}{dx}   =  -
      \frac{\lambda \langle \sigma v \rangle}{x^2 f_\phi(x)}~
      \left(2 \Delta_{\bar\chi} Y_{\bar\chi,{\rm eq}} +
       C \Delta_{\bar\chi} \right)\,.
\end{equation}
%
%
%
%
%
Repeating the same method as in \cite{Iminniyaz:2013cla}, we obtain
\begin{equation} \label{bardelta_solu}
      \Delta_{\bar\chi} \simeq \frac{2 x^2 f_\phi(x) P}
      {\lambda \langle \sigma v \rangle\,(C^2 + 4 P)}\,.
 \end{equation}
Where we used
$Y_{\bar\chi,{\rm eq}} = - C/2 + \sqrt{C^2/4 + P}$ which is
obtained by solving the Boltzmann equation Eq.(\ref{eq:Ycbar}) in equilibrium
state.
The freeze--out temperature $\bar{x}_F$ for $\bar{\chi}$ is determined by
using this solution.

On the other hand, for sufficiently low temperature, i.e. for $x > \bar x_F$,
we have $\Delta_{\bar\chi}\approx Y_{\bar\chi}\gg Y_{\bar\chi,{\rm eq}}$. The terms involving $dY_{\bar\chi,{\rm eq}}/dx$ and $Y_{\bar\chi,{\rm eq}}$ in the Boltzmann equation (\ref{eq:delta}) can be ignored, so that
\begin{equation} \label{eq:delta_late}
\frac{d \Delta_{\bar\chi}}{dx} = - \frac{\lambda \langle \sigma v
\rangle}{x^2 f_\phi(x)} \left( \Delta_{\bar\chi}^2 + C \Delta_{\bar\chi}
\right)\,.
\end{equation}
We integrate Eq.(\ref{eq:delta_late}) from $\bar{x}_F$ to $x_{\rm tr}$ where
the scalar--tensor theories applied and from $x_{\rm tr}$ to $\infty$ in which
period the standard cosmology is resumed. Again assuming
$\Delta_{\chi}(\bar{x}_F) \gg \Delta_{\chi}(\infty) $, we have
\begin{equation} \label{eq:barY_cross}
Y_{\bar\chi}(x \rightarrow \infty) =  \frac{C}
 { \exp \left[ \sqrt{\pi/45}\, C \, m_{\chi} M_{\rm P}\,
  I(\bar{x}_F,x_{\rm tr})   \right] -1}\,,
\end{equation}
where
\begin{eqnarray}
   I(\bar{x}_F, x_{\rm tr}) &=&
            \int^{x_{\rm tr}}_{\bar{x}_F} \frac{\sqrt{g_*} \langle \sigma v \rangle }
               {x^2~ f_\phi(x)}~ dx
              + \int^{\infty}_{x_{\rm tr}}
              \frac {\sqrt{g_*} \langle \sigma v \rangle}
             {x^2}~ dx \\
              &=& \sqrt{g_*}[A(\bar{x}_F)-A(x_{\rm tr})
              + B(x_{\rm tr}) ]\,,
\end{eqnarray}
with $\displaystyle A(x)=(m_\chi/{\rm GeV})^{0.82}(5.6938\times 10^{-5}\,a\,
x^{-1.82}+2.2048\times 10^{-4}\,b\, x^{-2.82})$, $B(x)=ax^{-1}+3bx^{-2}$.
The relic abundance for $\chi$ particle is obtained from Eq.(\ref{eq:c}) as
\begin{equation} \label{eq:Y_cross}
       Y_{\chi}(x \rightarrow \infty) =   \frac{C}
       { - \exp \left[-\sqrt{\pi/45}\, C \, m_{\chi} M_{\rm P}\,
       I(x_F, x_{\rm tr})  \right] + 1}\,,
\end{equation}
We note that only if $x_F = \bar x_F$, Eqs.(\ref{eq:barY_cross}) and
(\ref{eq:Y_cross}) are consistent each other with the constraint (\ref{eq:c}).
%
%
%
%

%
%
%
%

The final Dark Matter abundance can be expressed in a more canonical form as
usuall:
\begin{equation}
\Omega_\chi h^2 =\frac{m_\chi s_0 Y_{\chi}(x \to \infty)
h^2}{\rho_{\rm cr}}\,,
\end{equation}
with $s_0 = 2.9 \times 10^3~{\rm cm}^{-3}$ being the present entropy
density, and $\rho_{\rm cr} = 3 M_{\rm P}^2 H_0^2/(8\pi)$ the present
critical density, where $H_0$ is the Hubble constant.
The present relic density for Dark Matter is then predicted as
\begin{eqnarray} \label{omega}
 \Omega_{\rm DM}  h^2  =
    \frac{2.76 \times 10^8~ m_\chi ~C}
 { \exp \left[ \sqrt{\pi/45}\, C \, m_{\chi} M_{\rm P}\,
  I(\bar{x}_F,x_{\rm tr})   \right] -1}\,
      - \frac{2.76 \times 10^8~ m_\chi ~C}
       {  \exp \left[-\sqrt{\pi/45}\, C \, m_{\chi} M_{\rm P}\,
       I(x_F,x_{\rm tr} )  \right] - 1 }\,.
\end{eqnarray}
%

We need to fix the freeze--out temperature in order to calculate the final relic
density. When the deviation $\Delta_{\bar\chi} $ is of the same order as the
equilibrium value of $Y_{\bar\chi}$, the freeze--out occurs:
\begin{equation} \label{eq:xf1}
\xi Y_{\bar\chi,{\rm eq}}( \bar{x}_F) = \Delta_{\bar\chi}(
\bar{x}_F)\,,
\end{equation}
where  $\bar{x}_{F}$ is the freeze--out temperature which is obtained by
solving Eq.(\ref{eq:xf1})
and $\xi$ is a numerical constant of order unity, $\xi = \sqrt{2}-1$
\cite{standard-cos}. Using Eq.(\ref{bardelta_solu}), the freeze--out
temperature $\bar{x}_F$ is determined for $\bar{\chi}$ particles.

%
%


\section{Constraints on Parameter Space}
\setcounter{footnote}{0}

Nine-year Wilkinson Microwave Anisotropy Probe (WMAP) observations
give the  Dark Matter relic density as \cite{wmap}
\begin{eqnarray} \label{wmap}
  \Omega_{\rm DM} h^2 = 0.1138 \pm 0.0045\, ,
\end{eqnarray}
where $\Omega_{\rm DM}$ is the Dark Matter (DM) density in units of
the critical density, and $h = 0.738 \pm 0.024$ is the Hubble
constant in units of $ 100 {\rm km}/(\textrm{s}\cdot {\rm Mpc}) $. In this section, we use
this result to find the constraints on the parameter space of scalar--tensor
model. For asymmetric Dark Matter, the particle $\chi$ and
anti--particle $\bar{\chi}$ relic density account for total Dark Matter density,
\begin{equation} \label{add}
\Omega_{\rm DM}=\Omega_\chi + \Omega_{\bar{\chi}}\,.
\end{equation}
According to Eq.(\ref{wmap}), we use the following range,
\begin{equation} \label{range}
0.10 < \Omega_{\rm DM} h^2 < 0.12.
\end{equation}
\begin{figure}[h!]
  \begin{center}
    \hspace*{-0.5cm} \includegraphics*[width=9cm]{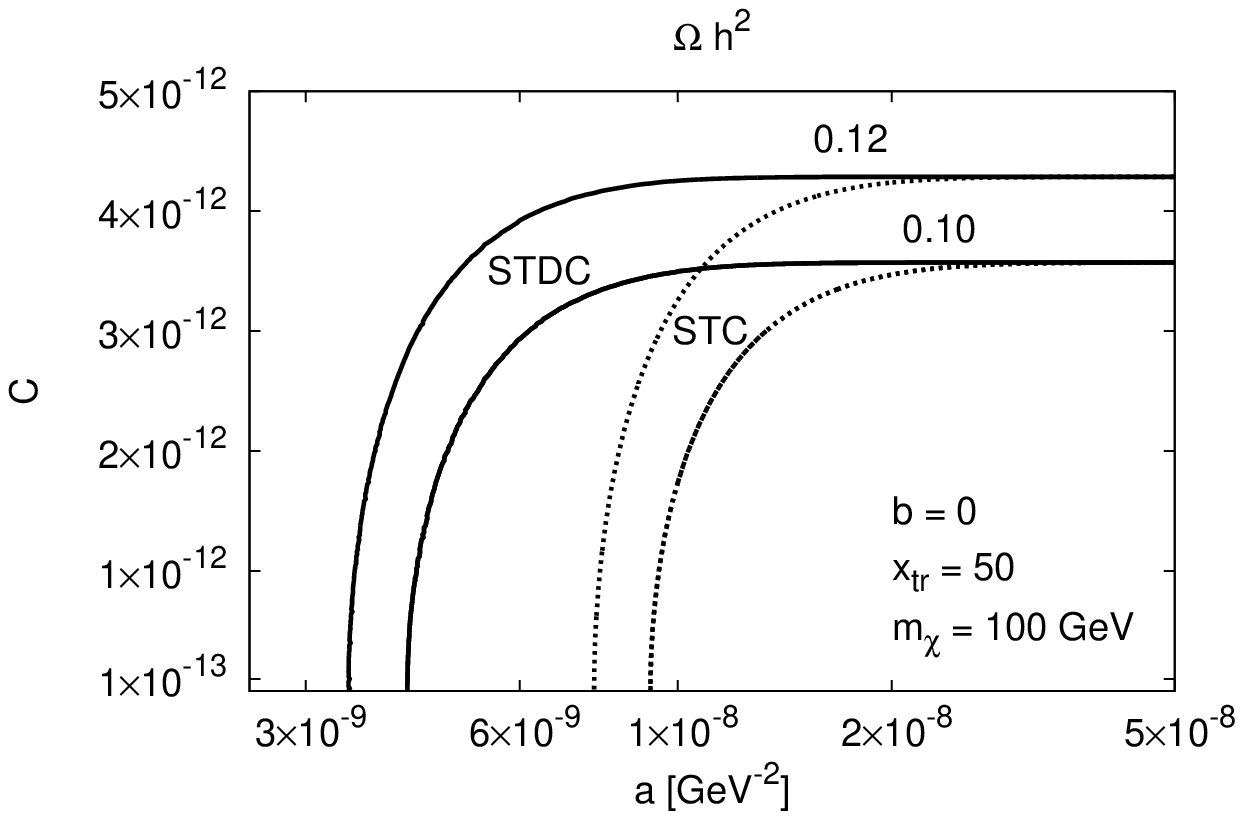}
    \put(-115,-12){(a)}
    \hspace*{-0.5cm} \includegraphics*[width=9cm]{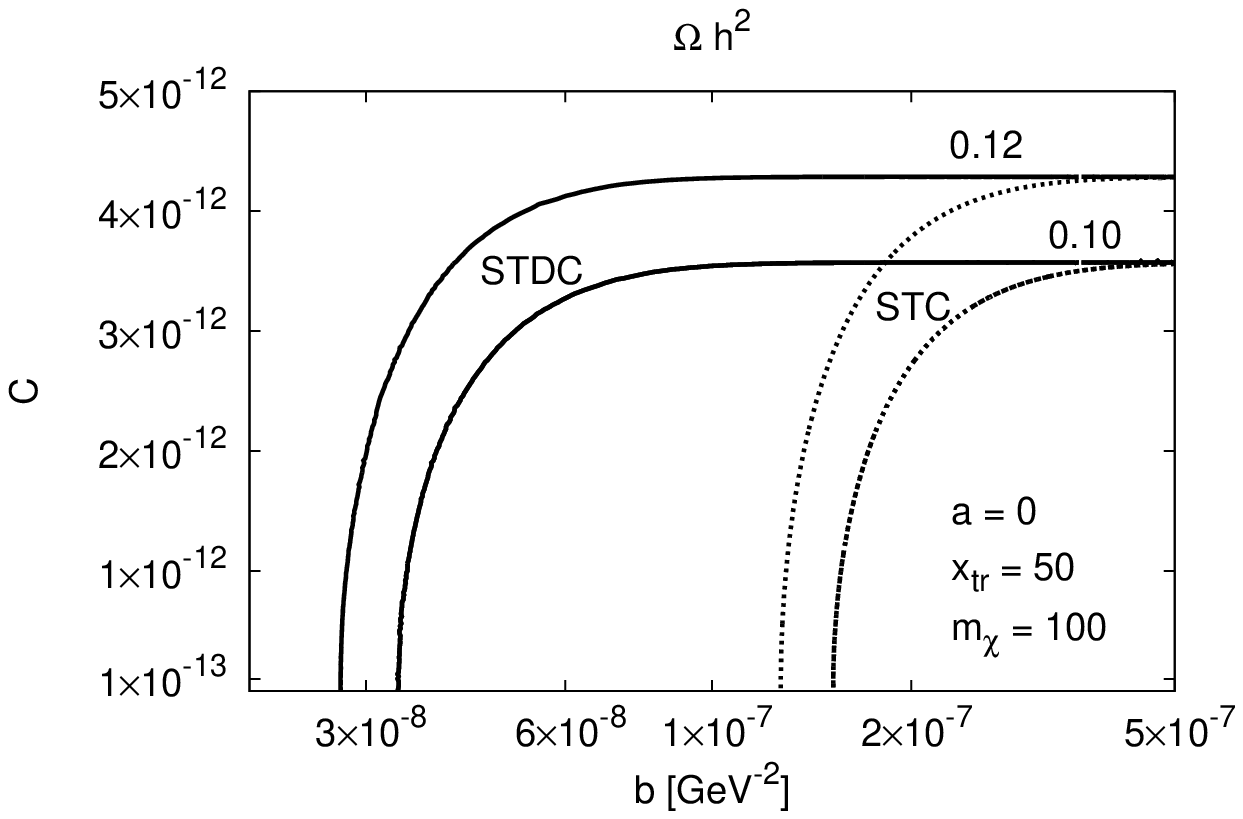}
    \put(-115,-12){(b)}
      \caption{\label{fig:d} \footnotesize
    The allowed region in the $(a, C)$ plane for $b=0$ (left) and
    in the $(b, C)$ plane for $a=0$ (right), when the Dark Matter density
    $\Omega_{\rm DM} h^2$ lies between 0.10 and 0.12. Here we take
    $m_\chi = 100$ GeV,
    $g_{\chi} = 2$ and $g_* = 90$, $x_{\rm tr}=50$. The solid lines are
    for the standard cosmological scenario (STDC), and the dotted lines are
    for scalar--tensor cosmology (STC). }
    \end{center}
\end{figure}

Fig.\ref{fig:d} shows the relation between the annihilation cross
section parameters $a, b$ and the asymmetry factor $C$ when the Dark
Matter relic density $\Omega_{\rm DM} h^2$ lies between 0.10 and 0.12.
Here we take $m_\chi = 100$ GeV, $g_{\chi} = 2$ and $g_* = 90$,
$x_{\rm tr}=50$. The solid lines are for the standard cosmological
scenario (STDC), the dotted lines are for the scalar--tensor
cosmology (STC). In the left frame (a) of Fig.\ref{fig:d}, it is
shown that for the values of s--wave annihilation cross sections
from $a = 3.5 \times 10^{-9}$ GeV$^{-2}$ to $a = 5.0 \times 10^{-8}$
GeV$^{-2}$, the observed Dark Matter abundance is obtained for the
range of $C$ from $0$ to $4.3 \times10^{-12}$ for
the standard cosmology. The initial cross--section is increased to $a
= 7.7 \times 10^{-9}$ GeV$^{-2}$ for the scalar--tensor cosmology to
obtain the observed Dark Matter abundance. In the right frame (b) of
Fig.\ref{fig:d}, for the asymmetry factor $C$ from $0$ to $4.4 \times10^{-12}$, 
one needs the p--wave annihilation cross sections from $b = 2.7 \times 10^{-8}$ GeV$^{-2}$
to $b = 5.0 \times 10^{-7}$ GeV$^{-2}$ to obtain the observed Dark
Matter abundance for standard cosmology. For the scalar-tensor
model, the corresponding p-wave cross section ranges from $b
= 1.3 \times 10^{-7}$ GeV$^{-2}$ to $b = 5.0 \times 10^{-7}$
GeV$^{-2}$. The cross section is increased evidently for the
scalar--tensor cosmology for both s--wave annihilation and p--wave
annihilation cross sections to obtain the observed Dark Matter
abundance. This is because in scalar--tensor model the interaction rate
is larger than the standard scenario. Asymmetric Dark Matter particles
need larger annihilation cross sections in order to fall in the observation
range.

\begin{figure}[t!]
  \begin{center}
    \hspace*{-0.5cm} \includegraphics*[width=8.7cm]{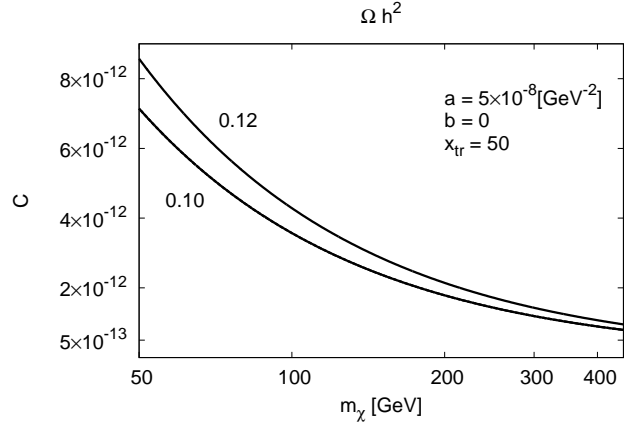}
    \caption{\label{fig:e1} \footnotesize
    The allowed region in the $(m_\chi, C)$ plane when the
    Dark Matter density $\Omega h^2$ lies between 0.10 and 0.12. Here we take
    $a = 5 \times 10^{-8}$ GeV$^{-2}$, $b=0$, $x_{\rm tr}=50$.}
   \end{center}
   \end{figure}
Fig.\ref{fig:e1} shows the allowed region in the $(m_\chi, C)$ plane when the
Dark Matter density $\Omega_{\rm DM} h^2$ lies between 0.10 and 0.12.
Here $a = 5 \times 10^{-8}$ GeV$^{-2}$, $b=0$, $x_{\rm tr}=50$. When the
mass of the Dark Matter increases from $50$ to 450 GeV, the asymmetry
factor $C$ decreaces from $C = 8.5 \times 10^{-12}$ to
$C = 8.0 \times 10^{-13}$ to obtain the
observed Dark Matter relic density. This can be understood from
Eq.(\ref{omega}). When $m_{\chi}$ increases, $C$ is
decreased in order to keep the observed Dark Matter abundance.
%
%
\begin{figure}[t!]
  \begin{center}
    \hspace*{-0.5cm} \includegraphics*[width=8.7cm]{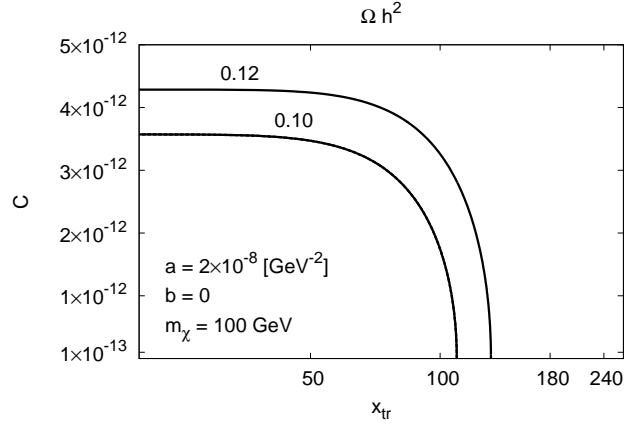}
    \caption{\label{fig:e} \footnotesize
    The allowed region in the $(x_{\rm tr}, C)$ plane for
$a = 2 \times 10^{-8}$ GeV$^{-2}$, $b=0$, $m_\chi=100$ GeV when the
    Dark Matter density $\Omega h^2$ lies between 0.10 and 0.12. }
    \end{center}
\end{figure}

The constraint of observed Dark Matter relic density on the
transition temperature in the scalar-tensor model is plotted in
Fig.\ref{fig:e}. The allowed region in the $(x_{\rm tr}, C)$ plane
for $a = 2 \times 10^{-8}$ GeV$^{-2}$, $b=0$, $m_\chi=100$ GeV is
shown when the Dark Matter relic density $\Omega_{\rm
DM} h^2$ lies between 0.10 and 0.12. When the inverse--scaled
transition temperature $x_{\rm tr}$ of the asymmetric Dark Matter in
the scalar--tensor model increases from $20$ to $130$, the asymmetry
factor $C$ decreases from $C = 4.3 \times 10^{-12}$ to $0$ to obtain
the observed Dark Matter relic density. This means that when the
inverse--scaled transition temperature increases the observed Dark Matter relic
density can be compensated by taking smaller asymmetry factor $C$.

%
\begin{figure}[t!]
  \begin{center}
    \hspace*{-0.5cm} \includegraphics*[width=8.7cm]{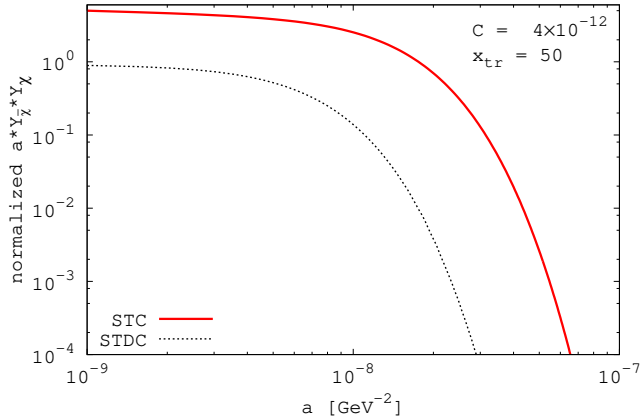}
    \caption{\label{fig:de} \footnotesize
    Annihilation rate for asymmetric Dark Matter in scalar--tensor cosmology
    and standard cosmology divided by the same quantity for symmetric Dark
    Matter ($C = 0$), as a function of $a$. The solid (red) line is
for the scalar--tensor cosmology and the dotted (black) line for the standard
cosmology. Here $b=0$, $m_\chi=100$ GeV and
$C = 4 \times 10^{-12}$, $x_{\rm tr} = 50$, $g_{\chi} = 2$, $g_* = 90$. }
    \end{center}
\end{figure}
The asymmetric Dark Matter can be detected by the direct detection.
Indirect detection for asymmetric Dark Matter in the standard
cosmological scenario, however, can not be used. This is because in
the standard scenario the symmetric part of the asymmetric Dark
Matter annihilated away with its anti--particles in the early
universe and only particles left in the present universe. For the
non--standard cosmological scenario like scalar--tensor cosmology,
there are still quite sizable amount of anti--particles left. Of
course the left amount of anti--particles mostly depends on the size of
annihilation cross section. Annihilation rate for asymmetric Dark Matter is
$\Gamma \propto \langle \sigma v \rangle Y_{\bar\chi} Y_{\chi} $.
Fig.\ref{fig:de} shows the annihilation rate for asymmetric Dark Matter in
scalar--tensor cosmology and standard cosmology which is divided by the same
quantity for the symmetric Dark Matter $C = 0$. The solid (red) line is for
scalar--tensor cosmology and the dotted (black) line is for standard
cosmology. We found the normalized annihilation rate
is increased for scalar--tensor cosmology comparing to the standard one. We
take an example, for $a = 10^{-8}$ GeV, the corresponding normalized
annihilation rate is $0.14$ for standard cosmology and $2.52$ for
scalar--tensor cosmology. This behavior makes the indirect
detection possible for scalar--tensor cosmology. For the larger cross
section as $ a \sim 1 \times 10^{-7}$ GeV$^{-2}$, the relic density of
anti--particles at the present day are almost depressed for scalar--tensor
cosmology too. This is indeed demonstrated in Fig.1 (c), (d). Therefore we
emphasize that the indirect detection signal constraints like
Fermi--LAT \cite{Ackermann:2011wa} can be used to nonstandard cosmological
scenarios like scalar--tensor model in conditionally.


\section{Summary and Conclusions}

In this paper we investigate the relic abundance of asymmetric WIMP
Dark Matter in scalar--tensor model. For asymmetric Dark Matter, the Dark Matter particles and anti--particles are distinct.
In the standard cosmological scenario, it is assumed that the particles were
in thermal equilibrium in the early universe and decoupled when they were
non--relativistic. The discussion of the relic density of asymmetric Dark
Matter in the standard cosmological scenario is extended to nonstandard
cosmological scenario. Specifically we discuss the scalar--tensor model. The
expansion rate is enhanced in scalar--tensor model and it leaves its imprint on the
relic density of asymmetric Dark Matter.

In this work, we give the exact numerical solutions and approximate analytical
solutions of the relic abundances for asymmetric Dark Matter in scalar--tensor
model. It is shown that the relic abundances for particles and anti--particles in scalar--tensor model are both greater than that of the standard cosmology. This is because the expansion rate is enhanced in scalar--tensor model. The particles and anti--particles freeze--out earlier than the standard case, leading to the increases of particle and anti--particle abundances. A significant feature of the relic abundances in scalar--tensor model is the sudden decrease at the transition temperatures $x_{\rm tr}$ in which the Hubble parameter returns to the standard cosmology.

We furthermore investigate the constraints on the transition temperature and
the asymmetry factor in the scalar--tensor model using
the observed Dark Matter abundance. The cross section is increased
approximately one order in scalar--tensor model in order to obtain the observed
Dark Matter abundance comparing to the standard cosmological scenario for
asymmetric Dark Matter. When the mass of the
asymmetric Dark Matter particles increases the asymmetry factor $C$ is
decreased to compensate the observed Dark Matter abundance.
If we take $a = 2 \times 10^{-8}$ GeV$^{-2}$, $b=0$, $m_\chi=100$ GeV, the inverse--scale transition temperature $x_{\rm tr}$ of the asymmetric Dark Matter in the scalar--tensor model increases from $20$ to $130$ as the asymmetry factor $C$ decreases from $C = 4.3 \times 10^{-12}$ to $0$ to obtain the observed Dark Matter relic density.

It is important to understand the relic abundance of asymmetric Dark
Matter in the early universe in nonstandard cosmological models. Our
work gives the constraints on the parameter space for the
scalar--tensor model in asymmetric Dark Matter case. Indirect
detection signal is possible for the nonstandard cosmological
scenarios due to the enhanced annihilation rate..

\section*{Acknowledgments}

The work of Shun-Zhi Wang is supported by Shanghai University of Engineering Science under grant
No.11XK11 and No.2011XZ10. The work of H. Iminniyaz is supported by the National Natural
Science Foundation of China (11365022, 11047009). Mamatrishat Mamat is supported by
National Natural Science Foundation of China (61366001).


\begin{thebibliography}{99}

\bibitem{Zwicky:1937zza}
  F.~Zwicky,  Astrophys. J. {\bf 86}, 217(1937).

\bibitem{Bertone:2004pz}
  G.~Bertone, D.~Hooper, and J.~Silk, Phys. Rept. {\bf 405},279 (2005).

\bibitem{wmap}
  G.~Hinshaw {\it et al.}  [WMAP Collaboration],
  Astrophys.\ J.\ Suppl.\ {\bf 208}, 19 (2013);
  C.~L.~Bennett {\it et al.}  [WMAP Collaboration],
  Astrophys.\ J.\ Suppl.\ {\bf 208}, 20 (2013).

\bibitem{PDG}
For a review, see K.~A.~Olive {\it et al.}  (PDG), Chin. Phys. {\bf
C38}, 090001 (2014)

\bibitem{review}
For a review, see G.~Jungman, M.~Kamionkowski, and K.~Griest,
Phys. Rep. {\bf 267}, 195 (1996)


\bibitem{adm-models}
S.~Nussinov, Phys.\ Lett.\ B {\bf 165}, 55 (1985);
K.~Griest and D.~Seckel.
Nucl. Phys. B {\bf 283}, 681 (1987);
S.~M.~ Barr, R.~S. ~Chivukula and E.~ Farhi,
 Phys.\ Lett.\ B {\bf 241}, 387 (1990);
R.~S.~Chivukula and T.~P.~Walker, Nucl.\ Phys.\ B {\bf 329}, 445 (1990);
D.~B.~Kaplan, Phys.\ Rev.\ Lett.\ {\bf 68}, 742 (1992);
D.~Hooper, J.~March-Russell and S.~M.~West,
 Phys.\ Lett.\  B {\bf 605}, 228 (2005) [arXiv:hep-ph/0410114];
JCAP {\bf 0901} (2009) 043 [arXiv:0811.4153v1 [hep-ph]];
H.~An, S.~L.~Chen, R.~N.~Mohapatra and Y.~Zhang,
  JHEP {\bf 1003}, 124 (2010) [arXiv:0911.4463 [hep-ph]];
T.~Cohen and K.~M.~Zurek,
  Phys.\ Rev.\ Lett.\  {\bf 104}, 101301 (2010) [arXiv:0909.2035 [hep-ph]].
D.~E.~Kaplan, M.~A.~Luty and K.~M.~Zurek,
  Phys.\ Rev.\  D {\bf 79}, 115016 (2009) [arXiv:0901.4117 [hep-ph]];
T.~Cohen, D.~J.~Phalen, A.~Pierce and K.~M.~Zurek,
Phys.\ Rev.\  D {\bf 82}, 056001 (2010) [arXiv:1005.1655 [hep-ph]];
J.~Shelton and K.~M.~Zurek,
Phys.\ Rev.\  D {\bf 82}, 123512 (2010) [arXiv:1008.1997 [hep-ph]];

\bibitem{frandsen}
A.~Belyaev, M.~T.~Frandsen, F.~Sannino and S.~Sarkar, Phys. Rev. D
{\bf 83}, 015007 (2011) [arXiv:1007.4839].

\bibitem{Gelmini:2006pw}
 G.~Giudice, E.~Kolb, and A.~Riotto
 Phys.\ Rev.\  D {\bf 64} (2001)  023508
  [arXiv:hep-ph/0005123];
  N.~Fornengo, A.~Riotto, and S.~Scopel
 Phys.\ Rev.\  D {\bf 67} (2003)  023514
 [arXiv:hep-ph/0208072];
 C.~Pallis
  Astorpart.\ Phys.\ {\bf 21} (2004)  689
 [arXiv:hep-ph/0402033];
  G.~Gelmini and P.~Gondolo,
  Phys.\ Rev.\  D {\bf 74} (2006)  023510
  [arXiv:hep-ph/0602230];
  M.~Drees, H.~Iminniyaz and M.~Kakizaki,
 Phys.\ Rev.\  D {\bf 73} (2006) 123502 [arXiv:hep-ph/0603165];
G.~Gelmini {\it et al}
 Phys.  Rev. D {\bf 74} (2006)083514
  [hep-ph/0605016].

\bibitem{Salati:2002md}
  P.~Salati,
  Phys.\ Lett.\ B {\bf 571}, 121 (2003)
  [astro-ph/0207396].

\bibitem{Kamionkowski}
M.~Kamionkowski and M.~S.~Turner,  Phys.\ Rev.\  D {\bf 42}, 3310 (1990).

\bibitem{Catena}
R.~Catena, N.~Fornengo, A.~Masiero, M.~Pietroni and F.~Rosati,
Phys. Rev.  D {\bf 70}, 063519 (2004) [arXiv:astro-ph/0403614].


\bibitem{Iminniyaz:2011yp}
  H.~Iminniyaz, M.~Drees and X.~Chen,
  JCAP {\bf 1107}, 003 (2011)
  [arXiv:1104.5548 [hep-ph]].

\bibitem{GSV}
M.~L.~Graesser, I.~M.~Shoemaker and L.~Vecchi, JHEP {\bf 1110}, 110 (2011)
  [arXiv:1103.2771 [hep-ph]].


\bibitem{Iminniyaz:2013cla}
  H.~Iminniyaz and X.~Chen,
  Astropart.\ Phys.\  {\bf 54}, 125 (2014)
  [arXiv:1308.0353 [hep-ph]].

\bibitem{Gelmini:2013awa}
  G.~B.~Gelmini, J.~H.~Huh and T.~Rehagen,
  arXiv:1304.3679 [hep-ph].  

\bibitem{Catena:2004ba}
  R.~Catena {\it et al.}
  N.~Fornengo, A.~Masiero, M.~Pietroni and F.~Rosati,
  Phys. Rev. D {\bf 70} (2004) 063519 [arXiv:astro-ph/0403614].


\bibitem{Santiago:1998ae}
  D.~Santiago, D.~Kalligas and R.~Wagoner,
  Phys.\ Rev.\  D {\bf 58} (1998) 124005.



\bibitem{standard-cos}
R.~J.~Scherrer and M.~S.~Turner, Phys. Rev. D {\bf 33}, 1585 (1986),
Erratum-ibid. D {\bf 34}, 3263 (1986).

\bibitem{Ackermann:2011wa}
 M.~Ackermann {\it et al.}  [Fermi-LAT Collaboration],
 Phys.\ Rev.\ Lett.\  {\bf 107}, 241302 (2011)  [arXiv:1108.3546 [astro-ph.HE]].



\end{thebibliography}
\end{document}